\documentclass[aip,reprint,pop]{revtex4-1}
\usepackage{graphicx}
\usepackage{subfigure}
\usepackage[table,dvipsnames]{xcolor}
\usepackage{longtable}
\usepackage{multirow}
\usepackage{epstopdf}
\usepackage{amsmath}
\usepackage{isomath}
\usepackage{amsfonts}
\usepackage{dcolumn}
\usepackage{bm}
\usepackage[colorlinks=true, linkcolor=blue]{hyperref}

\begin{document}

\title{A necessary condition for perpendicular electric field control in magnetized plasmas}
\author{Renaud Gueroult}
\affiliation{LAPLACE, Universit\'{e} de Toulouse, CNRS, INPT, UPS, 31062 Toulouse, France}
\author{Jean-Marcel Rax}
\affiliation{D\'epartement de Physique, Universit\'{e} de Paris XI - Ecole Polytechnique, 91128 Palaiseau, France}
\author{Nathaniel J. Fisch}
\affiliation{Department of Astrophysical Sciences, Princeton University, Princeton, New Jersey,
USA}

\begin{abstract}
The electrostatic model proposed by Poulos [Phys. Plasmas (2019), \textbf{26}, 022104] to describe the electric potential distribution across and along a magnetized plasma column is used to shed light onto the ability to control perpendicular electric fields. The effective electrical connection between facing end-electrodes is shown to be conditioned upon the smallness of a dimensionless parameter $\tau$ function of the plasma column aspect ratio and the square root of the conductivity ratio $\sigma_\perp/\sigma_{\parallel}$. The analysis of a selected set of past end-electrodes biasing experiments confirms that this parameter is small in experiments that have successfully demonstrated perpendicular electric field tailoring. On the other hand, this parameter is $\mathcal{O}(1)$ in experiments that failed to demonstrate control, pointing to an excessively large ion-neutral collision frequency. A better understanding of the various contributions to $\sigma_\perp$ is needed to gain further insights into end-biasing experimental results.  
\end{abstract}

\date{\today}

\maketitle

\section{Introduction}

Controlling plasma rotation holds promise for many important plasma applications. For magnetic confinement fusion, controlling rotation can provide the rotational transform ensuring particle confinement in toroidal geometry, but without the need for poloidal magnetic fields~\cite{Rax2017,Ochs2017b}. This alternative concept, referred to as the wave-driven rotating torus (WDRT), exhibits very natural advantages, including an improved efficiency compared to the classical RF current drive in tokamaks~\cite{Fisch1978} and a lower likelihood for plasma disruption. Controlling rotation also provides a means to affect the mass differential confinement properties that naturally arise as a result of centrifugal effects~\cite{Bonnevier1966,Lehnert1971}. These mass dependent phenomena then provide the basic components for designing plasma mass separation processes~\cite{Gueroult2018,Zweben2018} such as envisioned for nuclear waste cleanup~\cite{Gueroult2015}, spent nuclear fuel (SNF) reprocessing~\cite{Dolgolenko2017,Timofeev2014,Gueroult2014a,Vorona2015,Yuferov2017} and rare earth elements (REEs) recycling~\cite{Gueroult2018a}. 

Different schemes can in principle be employed to drive rotation. One possibility is external momentum input. Neutral beam injection is for instance routinely used to rotate plasma in tokamaks~\cite{Wade2007}. This technique might complicate plasma separation applications, since the additional beam ions will require removal.  Another possibility is to rely on rotating magnetic fields~\cite{Stepanov1958}. In this scheme the premise is that an infinite conductivity fluid plasma is dragged by the rotating magnetic field according to Alfv{\'e}n frozen in law~\cite{Alfven1942}. However, detailed analysis reveals that the single particle dynamics is much more complex than simple rotation~\cite{Soldatenkov1966,Rax2016}, and magnetic field penetration in the plasma is likely to only be effective for a limited range of plasma parameters~\cite{Milroy1999}. Yet another solution, instrumental both in the WDRT concept~\cite{Rax2017} and in plasma separation devices~\cite{Gueroult2019}, consists in employing crossed-field (or $\mathbf{E}\times\mathbf{B}$) configurations. Focusing in this paper on plasmas in which the magnetic field self-generated by plasma currents is negligible compared to the background field $\mathbf{B_0}$, controlling crossed-field driven rotation then boils down to controlling the electric potential profile in the direction perpendicular to the magnetic field. 

\begin{figure}
\begin{center}
\includegraphics[]{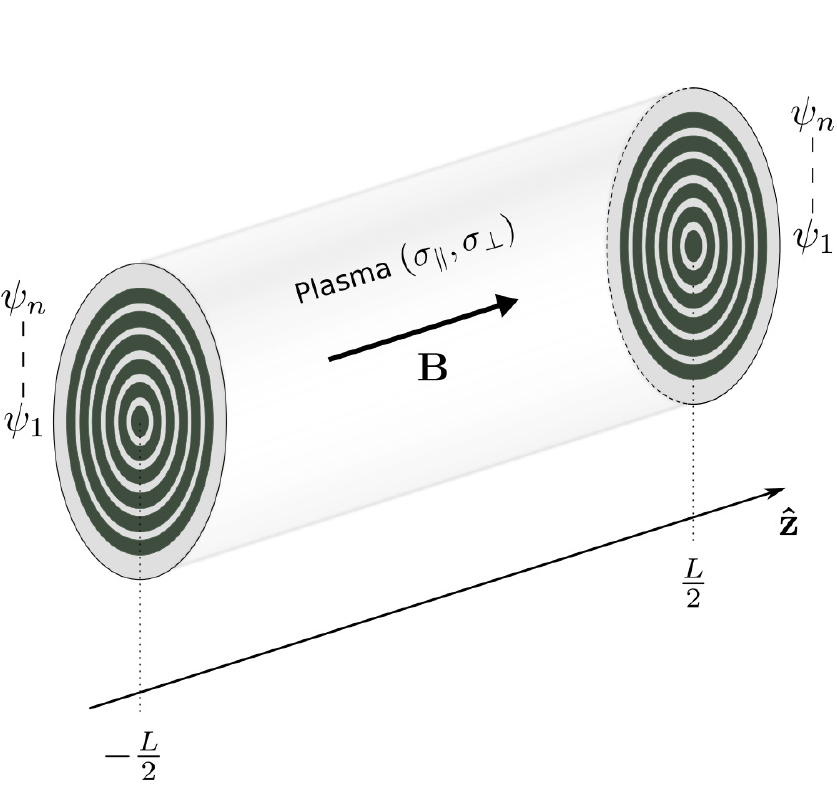}
\caption{Ring end-electrodes configuration. A magnetized plasma column terminates on two symmetrical sets of independently biased ring electrodes with potential ($\psi_1\cdots\psi_n$). }
\label{Fig:Rings}
\end{center}
\end{figure}

To examine this question, consider a plasma column immersed in a uniform background magnetic field $\mathbf{B}=B_0\mathbf{e_z}$. Because conductivity along field lines $\sigma_{\parallel}$ in a magnetized plasma is typically much greater than conductivity perpendicular to field lines $\sigma_{\perp}$, Lehnert suggested using a set of ring electrodes~\cite{Lehnert1970}, as illustrated in Fig.~\ref{Fig:Rings}. The proposition was that, by applying a suitable electric potential on each electrode, one can tailor the radial electric potential profile throughout the entire plasma column, with the potential of a given magnetic field line set by the electrode it intercepts. Even putting aside sheath effects at the electrodes, this \emph{magnetic line-tying} picture proposed by Lehnert~\cite{Lehnert1973} intrinsically assumes $\mu^{-1} = \sigma_{\parallel}/\sigma_{\perp}\rightarrow\infty$, or in other words that magnetic field lines are isopotential~\cite{Morozov2012}. Practically, assessing the efficiency of end-electrode biasing requires determining how the electric potential distributes itself along and across the plasma column for finite values of $\mu$. Lehnert studied how scattering with neutrals and viscosity constrain the practical parameter space for $B_0$ and plasma and neutral densities~\cite{Lehnert1973,Lehnert1974}. Bekhtenev~\emph{et al.}~\cite{Bekhtenev1978,Bekhtenev1980} then restated these constraints through the requirement for high conductivity between the plasma and the electrodes. Yet, a generic model describing potential distribution, and from there the effectiveness of end-electrodes biasing, for finite $\mu$ is still lacking. Note also that while we consider here this problem primarily with electrode biasing in mind, the question of potential distribution is equally critical for wave driven rotation~\cite{Fetterman2008,Fisch1992,Fetterman2009}.

Electrode biasing has been extensively used in experiments, primarily for the purpose of instability mitigation~\cite{Severn1991} and turbulence suppression through sheared plasma rotation~\cite{Schaffner2012}. Despite these differences in scope, these experimental studies provide indirect information on the effect of electrode biasing on the potential distribution, and hence offer opportunities to test new theoretical developments.

In this paper, we show how recent theoretical work by Poulos~\cite{Poulos2019} on the potential distribution in a magnetized plasma column provides, with the consideration of new boundary conditions, a necessary condition for effective perpendicular electric field control through end-electrodes biasing, and use this finding to offer a perspective on the results of number of past end-electrodes biasing experiments. In Sec.~\ref{Sec:SecII}, we review how the electric potential distribution in a homogeneous magnetized plasma column characterized by $\sigma_{\parallel}$ and $\sigma_{\perp}$ can be derived analytically, and identify a dimensionless parameter characterizing potential distribution across and along field lines. In Sec.~\ref{Sec:SecIII}, we provide rough estimates for this dimensionless parameter in various past end-electrodes biasing experiments, and analyse how it can shed light onto the effectiveness of potential tailoring in these experiments. Finally, the main findings are summarized in Sec.~\ref{Sec:SecIV}. 

\section{Electrostatic model}
\label{Sec:SecII}

In this section we largely follow the model proposed by Poulos~\cite{Poulos2019} to derive an analytical solution for the electric potential in a homogeneous magnetized plasma column. However, while Poulos~\cite{Poulos2019} considered an asymmetrical configuration corresponding to a discharge along magnetic field lines between an anode and a cathode, we focus here on a symmetrical configuration with symmetrical biased ring end-electrodes. The plasma column is assumed azimuthally symmetrical with uniform axial magnetic field $\textbf{B} = B_0 \hat{\textbf{z}}$, and characterized by uniform parallel and perpendicular conductivity $\sigma_{\parallel}$ and $\sigma_{\perp}$.

\subsection{General solution from Hankel transform}

Depending on whether fully ionized or partially ionized plasmas are considered, Ohm's law takes different forms in the literature. For fully ionized plasmas, an often used form is the \emph{generalized Ohm's law} given by Spitzer~\cite{Spitzer1962}, which in steady state reads
\begin{equation}
\bm{j} = \mathboldsans{\sigma}'[\mathbf{E}+\mathbf{v}\times\mathbf{B}]\label{Eq:ohm_center_mass}.
\end{equation}
Here $\bm{j}$ is the current density, $\mathbf{E}$ is the electric field and
\begin{equation}
\mathbf{v} = \mathbf{u}_i + \frac{m_e}{m_i+m_e}\mathbf{u}_e\label{Eq:v_center_mass}
\end{equation}
is the plasma center of mass velocity, with $\mathbf{u}_i$ and $\mathbf{u}_e$ the ion and electron velocity, respectively. In the study of ionospheric physics, an equivalent though different form~\cite{Kunkel1984,Kelley1989,Rax2015a,Richmond2016},
\begin{equation}
\bm{j} = \mathboldsans{\sigma}''[\mathbf{E}+\mathbf{v}_n\times\mathbf{B}]\label{Eq:ohm_wind}
\end{equation}
with $\mathbf{v}_n$ the neutral flow velocity, is often used to highlight the effect of neutrals. The replacement of $\mathbf{v}$ by $\mathbf{v}_n$ going from Eq.~(\ref{Eq:v_center_mass}) to Eq.~(\ref{Eq:ohm_wind}) stems from the fact that these two forms are written in different frames of reference (the plasma and neutrals rest frames, respectively). Note that, because of this substitution, the conductivity tensor $\bm{\sigma}$ which is typically a function of the collision frequencies also differs in each of this form, as indicated by the prime and double-prime notation. Note also that while Eq.~(\ref{Eq:ohm_wind}) was originally derived absent of accounting for Coulomb collisions, a similar equation can be obtained when both Coulomb collisions and collisions between charged particles and neutrals are modelled~\cite{Song2001}.

When the ion-neutral collision frequency $\nu_{in}$ is small compared to the ion gyro-frequency $\Omega_i$, one can typically assume $|\mathbf{v}_n|\ll|\mathbf{u}_i|\sim|\mathbf{v}|$ and $\mathbf{u}_i\sim \mathbf{E}\times\mathbf{B}/{B_0}^2$. This regime of weak ion-neutral coupling thus corresponds to a large ion-slip $\mathbf{u}_{is} = \mathbf{u}_i-\mathbf{v}_n \sim \mathbf{u}_{i}$~\cite{Kunkel1984,Sutton2006}. In these conditions $\mathbf{v}_n\times\mathbf{B}$ can be neglected in the bracketed term on the right hand side in Eq.~(\ref{Eq:ohm_wind}). Dropping for simplicity the double-prime notation, the simplified Ohm's law used by Poulos~\cite{Poulos2019}
\begin{subequations}
\label{Eq:Ohm}
\begin{align}
j_r & = -\sigma_{\perp} \frac{\partial \phi}{\partial r} \label{Eq:j_r} \\
j_z & = -\sigma_{\parallel} \frac{\partial \phi}{\partial z}
\end{align}
\end{subequations}
is then recovered, with $\phi$ the electric potential. As we will show in Sec.~\ref{Sec:SecIII}, the weak ion-neutral coupling condition $\nu_{in}\ll\Omega_i$ which ensures that Eq.~(\ref{Eq:Ohm}) is a valid approximation for Eq.~(\ref{Eq:ohm_wind}) is largely satisfied in the experiments considered in this study. In contrast, accounting for the $\mathbf{v_n}\times\mathbf{B}$ in Eq.~(\ref{Eq:ohm_wind}) will be essential for plasmas exhibiting a strong ion-neutral coupling (\emph{i.~e.} a small ion-slip $|\mathbf{u}_{is}|\ll|\mathbf{u}_{i}|$), such as in MHD converters and some high-pressure arc discharges~\cite{Kunkel1984,Sutton2006}.

Assuming quasi-neutrality, the charge continuity equation writes $\bm{\nabla}\cdot\bm{j} = 0$. Plugging in Eq.~(\ref{Eq:Ohm}) then leads to
\begin{equation}
\nabla_r^2 \phi + \frac{\sigma_{\parallel}}{\sigma_{\perp}} \nabla_z^2 \phi = 0.
\label{Eq:Potential}
\end{equation} 
Introducing the zero$^\textrm{th}$ order Hankel tranforrm~\cite{Jackson1970,Piessens1996} of $\phi$
\begin{equation}
\tilde{\phi}(k,z) = \mathcal{H}_0\{\phi(r,z)\} = \int_0^{\infty} r \phi(r,z) J_0(k r) dr,
\end{equation}
with $J_0$ the zero$^\textrm{th}$ order Bessel function of the first kind, Eq.~(\ref{Eq:Potential}) gives
\begin{equation}
-k^2 \tilde{\phi}(k,z) +\frac{\sigma_{\parallel}}{\sigma_{\perp}}\frac{\partial ^2}{\partial z^2} \tilde{\phi}(k,z) = 0.
\label{Eq:diff_Hankel}
\end{equation}
The general solution to Eq.~(\ref{Eq:diff_Hankel}) is then simply
\begin{equation}
\tilde{\phi}(k,z) = A^+(k)\exp[- k\sqrt{\mu}z] + A^-(k)\exp[ k\sqrt{\mu}z]
\end{equation}
with $\mu = \sigma_{\perp}/\sigma_{\parallel}$.

The potential can then be obtained from the inverse Hankel transform~\cite{Piessens1996}
\begin{align}
\phi(r,z) & = \mathcal{H}_0^{-1}\{\tilde{\phi}(k,z)\}\nonumber\\
 & = \int_0^{\infty}k \tilde{\phi}(k,z)J_0(k r) dk.
\end{align}
The functions $A^-(k)$ and $A^+(k)$ depend on the particular configuration and can be determined from boundary conditions. 

Before considering a particular electrode geometry, let us examine the effect of medium anisotropy on space charge density $\rho$. Combining $\bm{\nabla}\cdot\bm{j} = 0$ and  $\bm{\nabla}\cdot\mathbf{E} = \rho/\epsilon_0$, one gets
\begin{align}
\frac{\rho}{\epsilon_0} = & (\mu^{-1}-1) \frac{\partial^2\phi}{\partial z^2}\nonumber\\
 = &  -(1-\mu) \frac{1}{r}\frac{\partial}{\partial r}\left(r\frac{\partial\phi}{\partial r}\right).
 \label{Eq:space_charge}
\end{align}
In an anisotropic medium $\sigma_{\perp}\neq\sigma_{\parallel}$ \emph{i.~e.} $\mu\neq 1$, a space charge distribution $\rho(r,z)$ is therefore associated with the potential solution $\phi(r,z)$.

\subsection{Symmetrical ring electrodes}
\label{Sec:rings}

\begin{figure}
\includegraphics[]{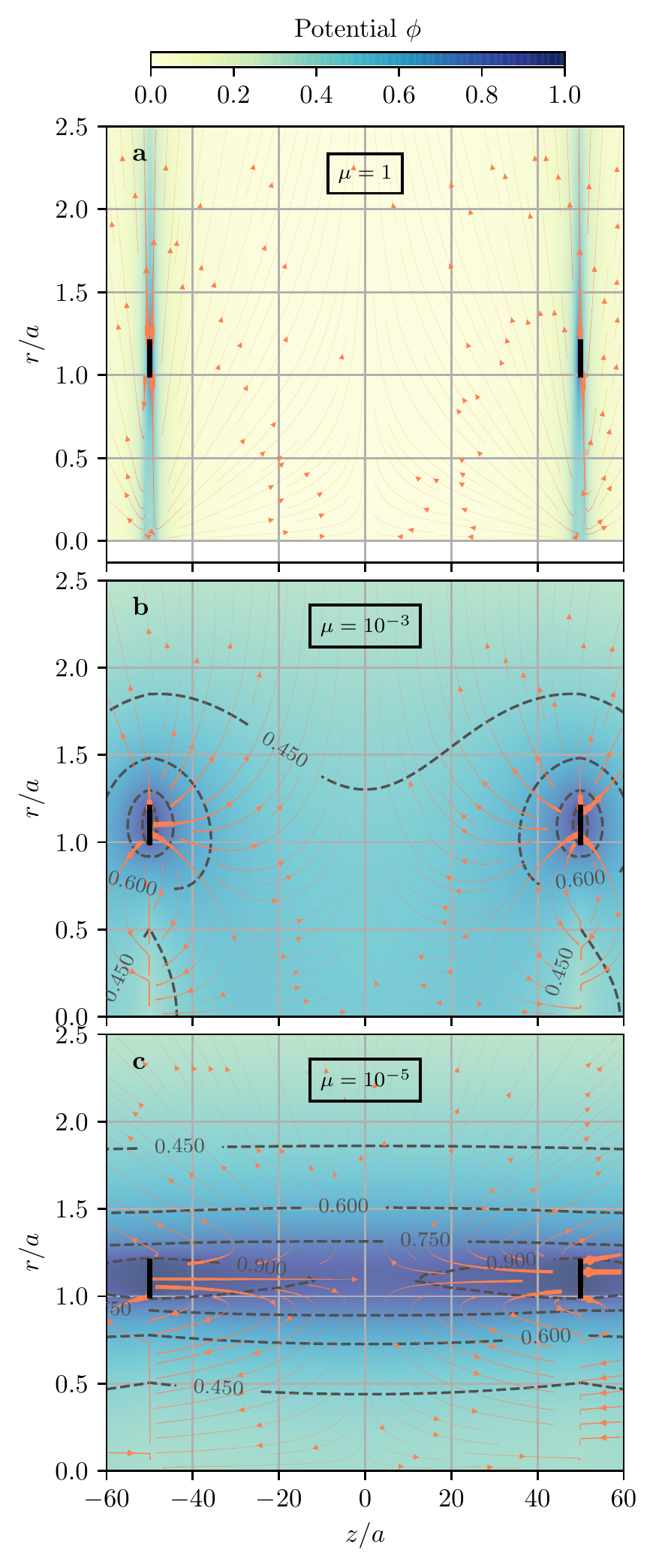}
\caption{Potential distribution $\phi(r,z)$ obtained from Eq.~(\ref{Eq:Solution_double_electrode}) for two identical ring electrodes (inner radius $a=1$, outer radius $b=1.2$) positioned at $z=\pm50a$ (in solid black) and different values of $\mu = \sigma_{\perp}/\sigma_{\parallel}$ corresponding to $\tau=100$, $\sqrt{10}$ and $1/\sqrt{10}$. The electrode potential $\psi_0=1$. Black dashed-lines are iso-potential lines. Coral solid lines are current density streamlines, with linewidth proportional to $|\bm{j}|$. }
\label{Fig:Fig2}
\end{figure}

Consider two identical ring electrodes (annuli with inner radius $a$ and outer radius $b$) positioned at $z=\pm L/2$. The boundary conditions then read
\begin{equation}
\phi(r,\pm L/2) = \psi_0 \quad \textrm{for} \quad a<r<b.
\label{Eq:BC_annular}
\end{equation}
One verifies that choosing 
\begin{equation}
\tilde{\phi}(k,z) = A(k)\frac{\exp[- k\sqrt{\mu}|z-L/2|]+\exp[- k\sqrt{\mu}|z+L/2|]}{1+\exp[-k\sqrt{\mu}L]}
\label{Eq:Hankel_Transform_dual_ring}
\end{equation}
with 
\begin{equation}
A(k) = \frac{2 \psi_0}{\pi k}\left[b j_0(kb)-a j_1(ka)\right]
\label{Eq:Ak}
\end{equation}
satisfies the Dirichlet condition on the ring electrodes pair. The first term on the right hand side of Eq.~(\ref{Eq:Ak}) corresponds to a disk shape parallel plate capacitor of radius $b$ at potential $\psi_0$, as studied by Atkinson \emph{et al.} in vacuum~\cite{Atkinson1983} (see the Appendix~\ref{Sec:appendixA} for the simple case of a single disk electrode). The second term corresponds to a hole of radius $a$ within the disks~\cite{Poulos2019}.  Expanding the denominator
\begin{equation}
\left(1+\exp[-k\sqrt{\mu}L]\right)^{-1} = \sum_{n=0}^{\infty} (-1)^n\exp[-nk\sqrt{\mu}L],
\end{equation}
Eq.~(\ref{Eq:Hankel_Transform_dual_ring}) rewrites
\begin{equation}
\tilde{\phi}(k,z) = A(k) \sum_{n=0}^{\infty} (-1)^n \left[\exp\left(-k\chi_n^-\right)+\exp\left(-k \chi_n^+\right)\right]
\end{equation}
with 
\begin{equation}
\chi_n^{\pm} = \sqrt{\mu}\left[nL+|z\pm L/2|\right].
\end{equation}
The solution for the potential is then
\begin{widetext}
\begin{equation}
\phi(r,z) =  \frac{2\psi_0}{\pi}\sum_{n=0}^{\infty} (-1)^n\int_0^{\infty} \left[b j_0(kb)-a j_1(ka)\right] J_0(k r) \left[\exp\left(-k\chi_n^-\right)+\exp\left(-k \chi_n^+\right)\right] dk.
\label{Eq:Solution_double_electrode}
\end{equation}
\end{widetext}
Recalling that~\cite{Gradshtein1980} 
\begin{subequations}
\begin{equation}
\int_0^{\infty} e^{-k \alpha}J_0(k \beta)\frac{\sin(k \gamma)}{k} dk = \arcsin\left(\frac{\gamma}{l_2}\right)
\label{Eq:Int_JO}
\end{equation}
and
\begin{equation}
\int_0^{\infty} e^{-k \alpha}J_1(k \beta)\frac{\sin(k \gamma)}{k} dk = \frac{\gamma-\sqrt{\gamma^2-{l_1}^2}}{\beta}
\end{equation}
\end{subequations}
with
\begin{subequations}
\begin{align}
l_1 &= \frac{1}{2}\left[\sqrt{\alpha^2+(\gamma+\beta)^2}-\sqrt{\alpha^2+(\gamma-\beta)^2}\right]\\
l_2 &= \frac{1}{2}\left[\sqrt{\alpha^2+(\beta+\gamma)^2}+\sqrt{\alpha^2+(\beta-\gamma)^2}\right],
\end{align}
\end{subequations}
and also that
\begin{subequations}
\begin{align}
\frac{\partial j_0(k a)}{\partial k} &= -a j_1(k a)\\
\frac{\partial J_0(k r)}{\partial k} &= -r J_1(k r),
\end{align}
\end{subequations}
Eq.~(\ref{Eq:Solution_double_electrode}) can be integrated by part in $k$-space to yield
\begin{align}
\phi(r,z) = & \frac{2\psi_0}{\pi}\sum_{n=0}^{\infty} (-1)^n \sum_{\pm} \arcsin\left(\frac{b}{\zeta_{b,n}^{\pm}}\right) \nonumber\\
 & \quad+ \frac{\chi_n^{\pm}}{a}\arcsin\left(\frac{a}{\zeta_{a,n}^{\pm}}\right) -\sqrt{1-\left(\frac{\xi_{a,n}^{\pm}}{a}\right)^2}
 \label{Eq:Solution_double_electrode_sum}
\end{align}
where we have introduced the variables
\begin{subequations}
\label{Eq:zeta_xi}
\begin{align}
\zeta_{\alpha,n}^{\pm} & = \frac{1}{2}\left[\sqrt{{\chi_n^{\pm}}^2+(r+\alpha)^2}+\sqrt{{\chi_n^{\pm}}^2+(r-\alpha)^2}\right]\\
\xi_{\alpha,n}^{\pm} & = \frac{1}{2}\left[\sqrt{{\chi_n^{\pm}}^2+(r+\alpha)^2}-\sqrt{{\chi_n^{\pm}}^2+(r-\alpha)^2}\right]
\end{align}
\end{subequations}
and the subscript $\alpha$ designates either $a$ or $b$. Numerical results show that  Eq.~(\ref{Eq:Solution_double_electrode_sum}) rapidly converges to Eq.~(\ref{Eq:Solution_double_electrode}). Quantitatively, the maximum relative error on the potential for the configuration corresponding to Fig.~\ref{Fig:Fig2}.b is below $10\%$ for $n_{max}\geq5$ and below $1\%$ for $n_{max}\geq45$, with $n_{max}$ the number of terms in the series Eq.~(\ref{Eq:Solution_double_electrode_sum}).


Take now a radial position centered with respect to the ring electrode $r_0 = (a+b)/2$. We see from Eq.~(\ref{Eq:zeta_xi}) that the characteristic distance over which the electrode potential $\psi_0$ is projected along the magnetic field line at $r=r_0$ is
\begin{equation}
\Delta = \frac{a}{\sqrt{\mu}}.
\end{equation}
In particular, we find that whether or not the facing ring electrodes are electrically connected is controlled by the dimensionless parameter 
\begin{equation}
\tau = \frac{L}{a}\sqrt{\mu},
\label{Eq:shape_factor}
\end{equation}
which is the inverse of the \emph{geometric factor} $s$ derived by Poulos~\cite{Poulos2019} in modelling potential partioning along field line in a magnetised plasma in the presence of an emissive cathode~\cite{Jin2019}. Electrical connection between facing electrodes is effective for $\tau\ll 1$, with the limit $\tau\rightarrow0$ corresponding to Lehnert's \emph{magnetic line-tying} picture~\cite{Lehnert1973}. On the other hand, the facing electrodes are not electrically connected for $\tau\gg 1$. The evolution of the potential distribution $\phi(r,z)$ as a function of $\tau$ is illustrated in Fig.~\ref{Fig:Fig2}. It is worth noting here that although the dimensionless parameter $\tau$ characterizing the electric potential distribution has been derived for facing annular ring electrodes at fixed potential, a similar criteria can be obtained when imposing a nearly parabolic potential profile at both ends of a plasma column (see Appendix~\ref{Sec:appendixB}). This suggests that $\sqrt{\mu}L/a$ is a robust parameter to describe the potential distribution along and across a plasma column.

Looking in Fig.~\ref{Fig:Fig2} at the streamlines of current density $\bm{j} = -\sigma_{\parallel}\partial \phi/\partial z \bm{\hat{z}} -\sigma_{\perp}\partial \phi/\partial r \bm{\hat{r}}$, one first notices that current emitted at the annular electrodes primarily flows to infinity. Indeed, since our model only imposes a fixed potential on the annular electrode (see Eq.~(\ref{Eq:BC_annular})) and does not include a reference electrode to which current could flow, regions at infinite distance from the electrodes where the potential vanishes act here as effective ground. A closer look further shows that, for the anisotropic cases (Fig.~\ref{Fig:Fig2}.b and Fig.~\ref{Fig:Fig2}.c), part of the current is drawn at $(0,\pm L/2)$, where local potential minima are found. This behaviour can be interpreted through the existence of a non-zero space charge $\rho$ when $\mu\neq1$. Indeed, as shown in Eq.~(\ref{Eq:space_charge}), $\nabla^2 \phi \neq 0$ for $\sigma_{\perp}\neq\sigma_{\parallel}$. 

\section{Revisiting results from end-electrodes biasing experiments}
\label{Sec:SecIII}

The analytical model derived in the previous section assumes a medium with uniform parallel and perpendicular conductivity, whereas density and temperature gradients, or magnetic field inhomogeneities, are expected to often invalidate this assumption in plasma experiments. In addition, this model neglects the axial variations of potential which are expected in the sheaths formed in front of the electrodes.  Despite these shortcomings, it is intriguing to confront this simple model with past end-electrodes biasing experiments. In particular, one wishes to know how the dimensionless parameter $\tau$ derived above possibly correlates with the effectiveness of end-electrodes biasing to control the potential throughout a plasma column.

However, even if neglecting gradients in plasma parameters, quantifying perpendicular conductivity $\sigma_{\perp}$ in a magnetized plasma represents a challenge. Beyond the classical collisionality driven conductivity, it is well established that instability and turbulence~\cite{Horton1999}, magnetic field fluctuations~\cite{Fenstermacher1983,Finn1992}, ion viscosity~\cite{Rozhansky2008,Kolmes2019} and particles sources and sinks~\cite{Kolmes2019} can all contribute to and enhance perpendicular conductivity. Recently, it has also been shown that the interplay between Coriolis, centrifugal, and collisional drag forces can lead to a non-linear and possibly non-local relation between the current density $\bm{j}$ and the non-homogeneous radial electric field $\mathbf{E}$ driving rotation in a magnetized plasma column~\cite{Rax2019,Kolmes2019}.

Notwithstanding the possibly important contribution of these additional phenomena, the analytic derivation suggests that a necessary but not sufficient condition for end-electrodes potential control is that $\tau\ll1$ at least for collisionally driven electric perpendicular conductivity. With that prediction in hand, we take another look at past experimental end-electrodes biasing results. To avoid further complications, we focus here on end-electrodes biasing experiments in linear geometry. In addition, and despite the promising results shown by emissive electrodes for electric field and rotation control~\cite{Jassby1972,Jin2019}, we further restrict our study to non-emissive electrodes, and consider a set of experiments surveyed in a previous study~\cite{Gueroult2019}. These experiments range from low density, low temperature plasmas used in the study of basic plasma phenomena~\cite{CarrollIII_RSI_1994,Tsushima1986} to high density, high temperature plasmas produced in magnetic confinement fusion devices~\cite{Severn1992,Tuszewski_PRL_2012} and thus offer the opportunity to test theoretical models across a wide range of plasma parameters. The machines on which were conducted these experiments are listed in Table~\ref{Table:Tab0}, and the particular plasma parameters used in each of these experiments are given in Table~\ref{Table:Tab1}.

\begin{table}
\begin{center}
\caption{Selected end-electrodes biasing experiments. }
\label{Table:Tab0}
\begin{tabular}{c c}
\hline
\hline
Experiment & Shorthand \\
 \hline
Q-machine, West Virginia Univ.	& Q-WVU\\
Large diameter helicon, Kyushu Univ. & H-KU\\
QT-Upgrade machine, Tohoku Univ. & M-TU\\
Gamma 10 mirror, Univ. Tsukuba & M-UT\\
LAPD afterglow, Univ. California Los Angeles &  L-UCLA\\
Phaedrus tandem mirror, Univ. Wisconsin & M-UW\\
Helcat, Univ. New Mexico	& H-UNM\\
PMFX, Princeton Plasma Physics Lab. & H-PPPL\\
C-2 device, Tri Alpha Energy & F-TAE\\
KMAX mirror, Univ. Sci. Tech. China & M-USTC\\
\hline
\hline
\end{tabular}
\end{center}
\end{table}

\begin{table*}
\begin{center}
\caption{Plasma parameters in selected end-electrodes biasing experiments: plasma density $n$, electron temperature $T_e$, ion temperature $T_i$, magnetic field $B$, neutral density $N$, ion atomic mass $M$, plasma column radius $a$, plasma length $L$ and typical applied bias $\psi_0$ with respect to machine ground. For mirror experiments (QT-Upgrade, Gamma 10 and Phaedrus), the parameters are taken in the middle of the central cell. For the field reverse configuration (C-2), parameters are taken at the edge.}
\label{Table:Tab1}
\begin{tabular}{c c c c c c c c c c c}
\hline
\hline
 & $n$ [$\times 10^{12}$~cm$^{-3}$] & $T_e$ [eV] & $T_i$ [eV] & $B$ [kG] & $N$ [$\times 10^{12}$~cm$^{-3}$] & $M$ [amu] & $a$ [cm] & $L$ [m] & $\psi_0$ [V] & Refs.\\
 \hline
Q-WVU & $0.001$ & $0.2$ & $0.2$ & $1.5$ & $0.099$ & $39$ & $3$ & $0.8$ & $-6$ to $18$ & \cite{Koepke_PRL_1994, CarrollIII_RSI_1994}\\
H-KU & $0.01$ & $4.5$ & $0.2$ & $1$ & $5.3$ & $40,131$ & $22$ & $1.7$& $0$ to $250$ & \cite{Shinohara2007}\\
M-TU & $0.3$ & $5$ & $0.8 \pm 0.2$ & $1.5$ & $1.6$ & $40$ & $8.5$ & $1.4$& $-20$ to $20$ & \cite{Tsushima1986, Tsushima1991}\\
M-UT & $0.75 \pm 0.25$ & $90 \pm 30$ & $650 \pm 150$ & $4$ & $0.033$ & $1$ & $20$ & $27$& $-2000$ to $750$ & \cite{Inutake_PRL_1985, Mase_NF_1991, Saito_PoP_1995}\\
L-UCLA & $0.2$ & $1$ & $1$ & $1$ & $3.3$ & $4$ & $15$ & $13$& $0$ to $150$ & \cite{Koepke_PPCF_2008,Finnegan_PPCF_2008,Gekelman2016}\\
M-UW & $2$ & $17.5 \pm 2.5$ & $30$ & $0.6$ & $5 \pm 5$ & $1$ & $17$ & $10$ & $-60$ to $60$ & \cite{Severn1992}\\
H-UNM & $10$ & $4 \pm 1$ & $0.2$ & $0.44$ & $66 \pm 33$ & $40$ & $6$ & $3$& $0$ to $40$ & \cite{Lynn2009, Gilmore2009}\\
H-PPPL & $10$ & $5 \pm 2$ & $0.2$ & $0.7 \pm 0.3$ & $165 \pm 33$ & $40,84$ & $6$ & $0.4$ & $-25$ to $15$ & \cite{Gueroult2016a}\\
F-TAE & $40$ & $200$ & $500$ & $1$ & $0.05 \pm 0.05$ & $2$ & $50$ & $15$& $-1000$ & \cite{Tuszewski_PRL_2012, Gupta_RSI_2012}\\
M-USTC & $0.10 \pm 0.05$ & $7 \pm 3$ & $0.2$ & $0.3 \pm 0.1$ & $25$ & $1$ & $12$ & $9$& $-30$ to $350$ & \cite{Zhang_FST_2015,Shi_PoP_2019}\\
\hline
\hline
\end{tabular}
\end{center}
\end{table*}

\subsection{Collisionallity driven perpendicular conductivity}

\begin{figure}
\begin{center}
\includegraphics[]{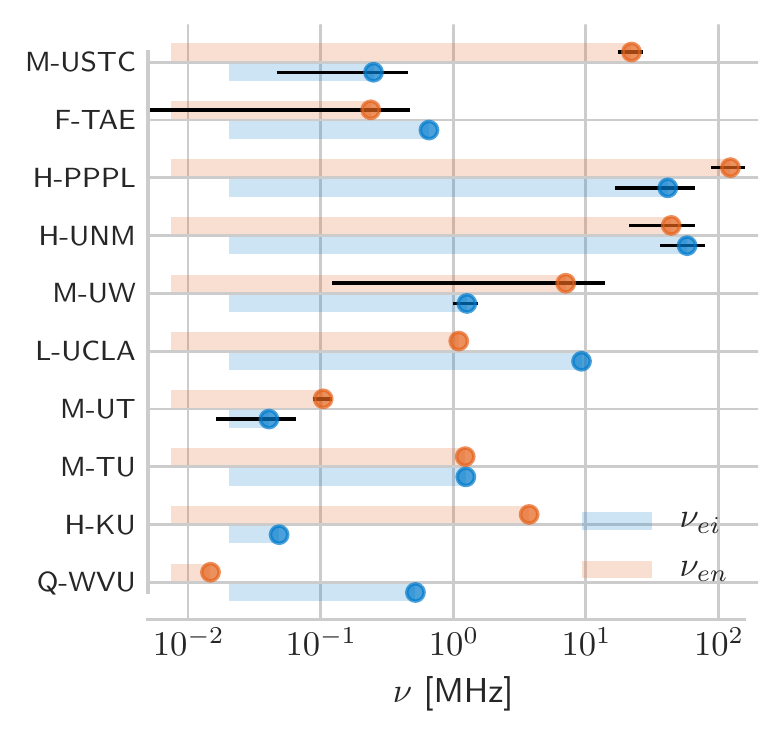}
\caption{Electron-ion and electron-neutral collision frequencies $\nu_{ei}$ and $\nu_{en}$ for selected end-electrodes experiments (see Table~\ref{Table:Tab0}). Error bars stem primarily from uncertainties on the neutral density $N$, and to a lower extent from uncertainties on the electron temperature $T_e$. }
\label{Fig:Electron_collision_freq}
\end{center}
\end{figure}

As shown in Table~\ref{Table:Tab1}, experiments range from fully ionized to partially ionized plasmas, and cover a large range of electron temperature. As a result, the ordering between electron-ion and electron-neutral collision frequencies $\nu_{ei}$ and $\nu_{en}$ is not consistent across these experiments. The parallel conductivity
\begin{equation}
\sigma_{\parallel} = \frac{n e^2}{m_e(\nu_{ei}+\nu_{en})}
\end{equation}
with $e$ and $m_e$ the electron charge and mass and $n$ the plasma density can either be governed by neutrals or ions, or both. The respective contributions to parallel conductivity are estimated using the order of magnitude estimate for electron-neutral collision
\begin{equation}
\nu_{en} = \sigma_0 N \sqrt{\frac{8 k_b T_e}{\pi m_e}}
\end{equation}
with $T_e$ the electron temperature, $N$ the neutral density and $\sigma_0$ a fiducial cross section taken to be $5\times10^{-15}$~cm$^{-2}$, and the electron-ion collision frequency
\begin{equation}
\nu_{ei} = \frac{nZ^2e^4\Lambda}{6\sqrt{2}\pi^{3/2}{\epsilon_0}^2\sqrt{m_e}(k_b T_e)^{3/2}}
 \end{equation}
with $Z$ the ion charge state and $\Lambda$ the Coulomb logarithm. Note that these are order of magnitude estimates, and that a detailed modelling will require accounting for the temperature dependence of $\sigma_0$. In addition, with the exception of F-TAE and M-UW, the neutral densities $N$ given in Table~\ref{Table:Tab1} correspond to gas fill pressure and not actual measurements, which adds further uncertainty to the collision frequency estimates. Quantitatively, 
\begin{equation}
\nu_{en} = 0.3\times N_{12} \sqrt{T_{e_{\textrm{eV}}}}~\textrm{MHz}
\end{equation}
and, using the canonical value $\Lambda=16$,
\begin{equation}
\nu_{ei} =  46\times n_{12} Z^2 {T_{e_\textrm{eV}}}^{-3/2}~\textrm{MHz}
\end{equation}
with $n_{12}$ and $N_{12}$ the plasma and neutral densities in $10^{12}$~cm$^{-3}$ and $T_{e_{\textrm{eV}}}$ the electron temperature in eV. The crossover point where $\nu_{ei} =\nu_{en}$ is found to occur for a plasma to neutral density ratio
\begin{equation}
\left.\frac{n}{N}\right |_{c} = 6.5\times 10^{-3} ~Z^{-2}{T_{e_\textrm{eV}}}^{2},
\label{Eq:cross_over}
\end{equation}
which is about $0.15$ and $65$ for $Z=1$ and $T_{e_\textrm{eV}} = 5$ and $100$, respectively. The large spread of collisional regimes is illustrated in Fig.~\ref{Fig:Electron_collision_freq}.
 
As shown in Table~\ref{Table:Tab2}, the ion thermal gyro-radius $\rho_{th,i}= v_{th,i}/\Omega_i$, with $\Omega_i$ and $v_{th,i}$ the ion gyro-frequency and ion thermal speed, is typically $0.1-1$~cm, and the ion thermal gyro-radius to perpendicular length scale ratio $\rho_{th,i}/a$ is smaller than $1$ in these experiments. Ions can thus be considered magnetized. We also verify in table~\ref{Table:Tab2} that $\nu_{in}/\Omega_i\ll 1$ for most experiments, and that at least $\nu_{in}/\Omega_i\leq 1$, supporting the assumption of a weak coupling between ions and neutrals made in our derivation of Eq.~(\ref{Eq:Ohm}). Focusing here on collisionality driven perpendicular conductivity, $\sigma_{\perp}$ is driven by ion-neutral collisions and characterized by Pedersen conductivity~\cite{Richmond2016}
\begin{equation}
\sigma_P = \frac{\displaystyle  n e^2\nu_{in}}{\displaystyle m_i({\nu_{in}}^2+{\Omega_i}^2)},
\label{Eq:sigma_perdersen}
\end{equation}
with $\nu_{in}$ the ion-neutral collision frequency. Using similarly the simple estimate
\begin{equation}
\nu_{in} = \sigma_0 N \sqrt{\frac{8 k_b T_i}{\pi m_i}},
\end{equation}
one finds that, for the experiments considered here, ${\nu_{in}}^2$ can be neglected in front of ${\Omega_i}^2$ in the denominator of Eq.~(\ref{Eq:sigma_perdersen}), and $\sigma_P\sim n e \nu_{in}B^{-1}\Omega_i^{-1}$. Quantitatively, 
\begin{equation}
\sigma_P\sim 1.3\times 10^{-3} n_{12}N_{12}\frac{\sqrt{M T_{i_{\textrm{eV}}}}}{Z{B_{\textrm{kG}}}^2}~\textrm{Ohm}^{-1}.\textrm{m}^{-1}
\label{Eq:sigma_perdersen_quant}
\end{equation}
with $T_{i_{\textrm{eV}}}$ the ion temperature in eV, $M$ the ion atomic mass and $B_{\textrm{kG}}$ the magnetic field in kG.

\begin{figure}
\begin{center}
\includegraphics[]{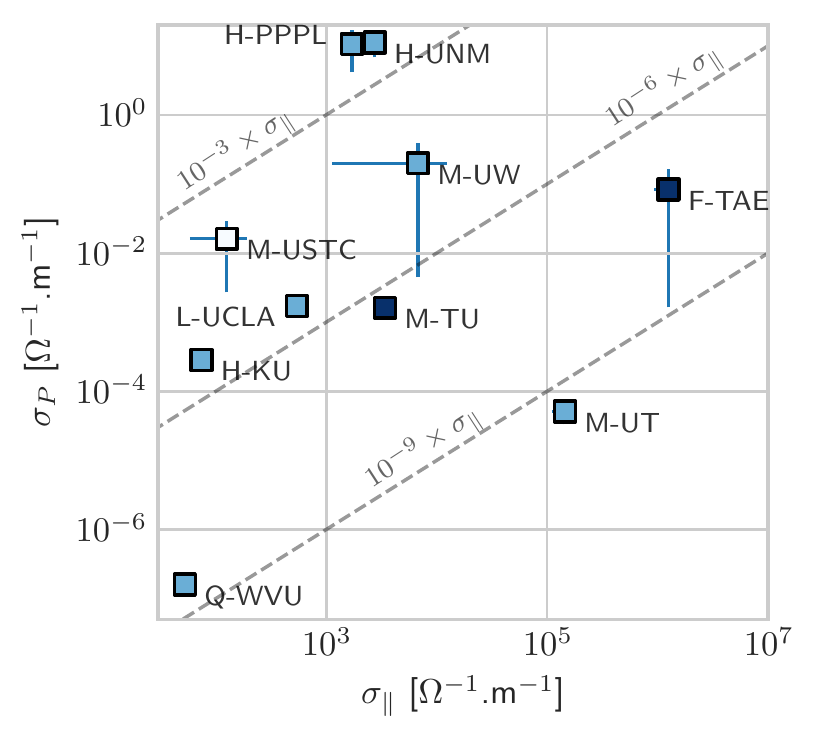}
\caption{Distribution of selected experiments (see Table~\ref{Table:Tab0}) in $(\sigma_{\parallel},\sigma_P)$ space.}
\label{Fig:ScatterSigmas}
\end{center}
\end{figure}

The distribution in $(\sigma_{\parallel},\sigma_P)$ space of the selected experiments obtained using these simple estimates is shown in Fig.~\ref{Fig:ScatterSigmas}. The helicon experiments H-PPPL and H-UNM feature the largest $\sigma_P$ ($\sim 10~\Omega^{-1}.$m$^{-1}$) due to high plasma density and comparatively high operating pressure. On the other hand, these plasma and neutral densities, combined with moderate electron temperature, lead to intermediate $\sigma_{\parallel}$ ($\sim 10^3~\Omega^{-1}.$m$^{-1}$). For similar temperatures, a drop by two orders of magnitude of both plasma and neutral densities leaves $\sigma_{\parallel}$ mostly unaltered, but decreases $\sigma_P$ by 4 orders of magnitude ($\sim 10^{-3}~\Omega^{-1}.$m$^{-1}$) for the electron cyclotron resonance (ECR) mirror experiment M-TU. Similar $\sigma_P$ can be obtained at lower $\sigma_{\parallel}$ from a combined increase and decrease of $N$ and $n$, as found in the radio-frequency (RF) plasma experiment H-KU, or from a decrease of $T_e$, as found in the hot-cathode experiment L-UCLA. Alternatively, larger $\sigma_{\parallel}$ ($\sim 10^6~\Omega^{-1}.$m$^{-1}$) are found for the higher $T_e$ plasmas in the high power mirror experiment M-UT and the field reverse configuration (FRC) experiment F-TAE. Finally, both low $\sigma_{\parallel}$ and low $\sigma_P$ ($\sim 10^{-7}$ and $10~\Omega^{-1}.$m$^{-1}$) are estimated for the low temperature and low density plasma of the Q-machine Q-WVU. 

\begin{figure}
\begin{center}
\includegraphics[]{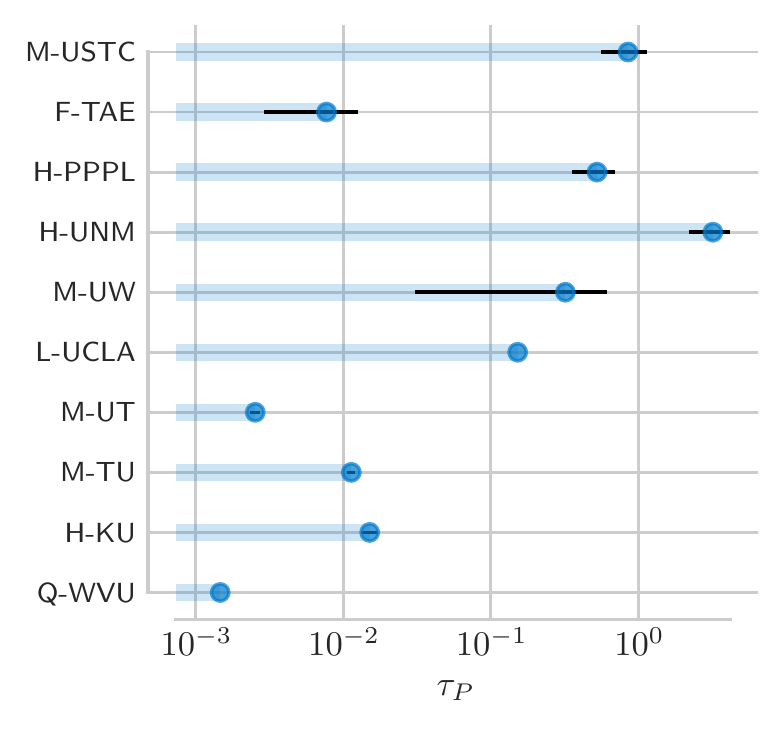}
\caption{Estimated dimensionless parameter $\tau_P=L/a\sqrt{\sigma_P/\sigma_{\parallel}}$ in selected end-electrodes biasing experiments (see Table~\ref{Table:Tab0}). }
\label{Fig:Adim_param_s}
\end{center}
\end{figure}

Plugging in the plasma radius $a$ and length $L$ yields the dimensionless parameter
\begin{equation}
\tau_P = \frac{L}{a}\sqrt{\frac{\sigma_P}{\sigma_{\parallel}}}
\end{equation}
computed for the Pedersen perpendicular conductivity $\sigma_P$ for each of the selected experiments. Quantitatively, two formulas can be derived for electron-neutral collisions and electron-ion collisions dominated parallel conductivity, with respectively
\begin{subequations}
\begin{multline}
\tau_P = 1.2\times10^{-4}\frac{L}{a}\frac{N_{12}}{B_{\textrm{kG}}\sqrt{Z}}\left(M{T_{i_{\textrm{eV}}}}T_{e_{\textrm{eV}}}\right)^{1/4}\\
\textrm{if}\quad \frac{n_{12}}{N_{12}}\left(\frac{T_{e_{\textrm{eV}}}}{Z}\right)^2\gg1.5\times10^2
\label{Eq:s_p_en}
\end{multline}
and 
\begin{multline}
\tau_P = 1.5\times 10^{-3} \frac{L}{a}\frac{M^{1/4}{T_{i_{\textrm{eV}}}}^{1/4}}{B_{\textrm{kG}}}\sqrt{\frac{\displaystyle n_{12}N_{12}Z}{\displaystyle{T_{e_{\textrm{eV}}}}^{3/2}}}\\
\textrm{if}\quad \frac{n_{12}}{N_{12}}\left(\frac{T_{e_{\textrm{eV}}}}{Z}\right)^2\ll1.5\times10^2.
\label{Eq:s_p_in}
\end{multline}

\end{subequations}
The results are plotted in Fig.~\ref{Fig:Adim_param_s}. Although the conductivities used here are at best very crude estimates and $\sigma_{\perp}$ accounts only for ion-neutral collisions, Fig.~\ref{Fig:Adim_param_s} already reveals interesting trends. A group of experiments formed by H-PPPL, M-USTC and H-UNM is found to have $\tau_P=\mathcal{O}(1)$, which, according to the simple model derived in this study, prohibits effective connection between end-electrodes. Interestingly, this is consistent with experimental measurements showing large potential variation along the machine axis in M-USTC~\cite{Zhang_FST_2015} and limited potential control. At the other end of the spectrum, one finds another group of experiments for which $10^{-3}\leq \tau_P\leq 10^{-2}$. Barring other possible contributions to $\sigma_{\perp}$, this in principle allows for end-electrodes potential control. Among these is found M-TU, for which electric connectivity between end-electrodes has been demonstrated, and radial electric fields with both polarities have been produced~\cite{Tsushima1986}.

The partial correlation observed experimentally between $\tau$ and end-electrode electric field control is encouraging. It suggests that this simple scaling can indeed be used to identify a priori plasma regimes where end-electrode biasing is ineffective in controlling the perpendicular electric field. However, it is still insufficient to predict regimes where end-electrode biasing should be effective. To enable such predictive capabilities, the model will have to be extended to include possible additional contributions to conductivity. It should also capture possible coupling effects between these different contributions. In addition, spatial gradients of plasma parameters are also expected, under certain conditions, to play a role and should be included in a global perpendicular conductivity model. Although such an extensive analysis goes well beyond the scope of this study, we briefly illustrate this task in the next paragraph by considering the possible contribution of inertia to perpendicular conductivity. Finally, detailed comparison with experiments will require supplementing conductivity models with sheath models. Indeed, even if assuming $\tau\rightarrow 0$, the electric field imposed on ring electrodes (see Fig.~\ref{Fig:Rings}) will be identically recovered in the plasma column only if the voltage drop across the sheath does not vary significantly along the machine radius. For a classical voltage drop $\Delta \phi$ across the sheath of a few $T_e$, this in turn sets an upper limit on how large the electron temperature gradient can be for a given target radial electric field. 

\subsection{Contribution of inertia in rotating plasmas}
\label{Sec:IIIb}

In a rotating plasma an additional contribution to perpendicular conductivity stems from inertia~\cite{Rax2019,Kolmes2019}. For a fully ionized plasma with limited shear, it writes~\cite{Rax2019} 
\begin{equation}
\sigma_{\Omega} = 2\frac{\displaystyle n Z e}{\displaystyle B}\frac{ \nu_{ie}\Omega^2}{\displaystyle {\Omega_i}^3} \sim 2Z\frac{\nu_{ie}}{\nu_{in}}\frac{\Omega^2}{{\Omega_i}^2} \sigma_P
\label{Eq:sigma_omega}
\end{equation}
and one shows that
\begin{multline}
\sigma_{\Omega} = 9.1\times10^{-7} M^2 {n_{12}}^2 {T_{e_\textrm{eV}}}^{-3/2} {B_{\textrm{kG}}}^{-6} {E_{\textrm{cm}}}^2{r_{\textrm{cm}}}^{-2}~\\\textrm{Ohm}^{-1}.\textrm{m}^{-1}
\end{multline}
with $E_{\textrm{cm}}$ the electric field in V.cm$^{-1}$, $r_{\textrm{cm}}$ the position in cm and where the lowest order expansion of the angular frequency $\Omega= E/(rB)$ has been assumed for simplicity.

\begin{figure}
\begin{center}
\includegraphics[]{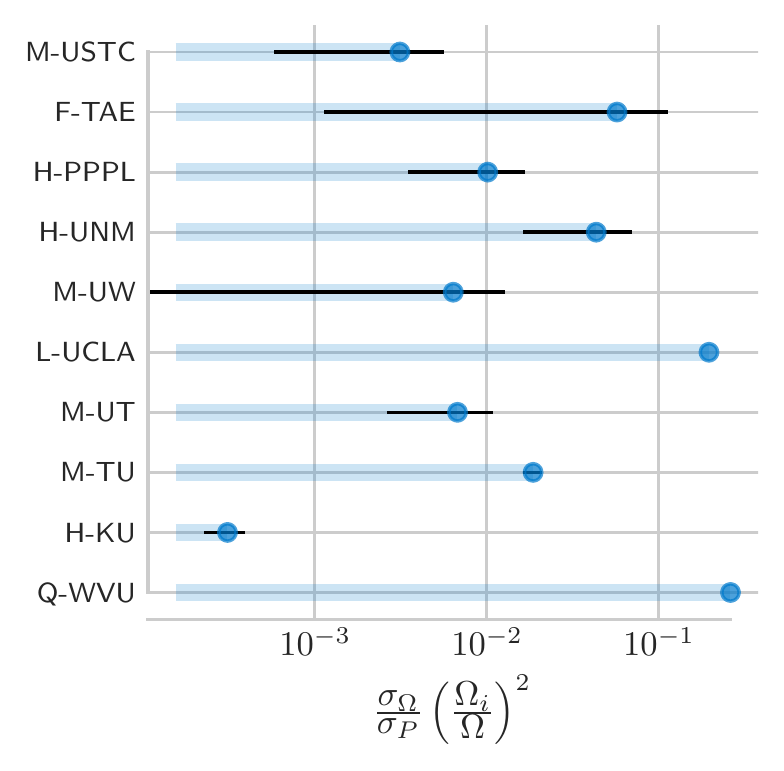}
\caption{Relative contribution of inertia effects in selected end-electrodes biasing experiments (see Table~\ref{Table:Tab0}). }
\label{Fig:Normalized_inertia_contribution}
\end{center}
\end{figure}

Assessing the relative contribution of inertia in front of collisionality driven conductivity requires estimating the radial electric field in these experiments. Unfortunately, potential radial profiles are not available for all selected biasing experiments and operating conditions. Short of information on $E$, a reasonable assumption to infer the relative importance of $\sigma_{\Omega}$ compared to $\sigma_{P}$ is to take $\Omega/\Omega_i\ll1$. This condition is indeed verified in most experiments since the large free energy content in the azimuthal drift motion lead to instabilities when $\Omega/\Omega_i=\mathcal{O}(1)$~\cite{Gueroult2017b}. Plugging in data given in Table~\ref{Table:Tab1} shows that the ratio 
\begin{equation}
\frac{\displaystyle \sigma_{\Omega}}{\displaystyle \sigma_{P}} \left(\frac{\displaystyle \Omega_i}{\displaystyle \Omega}\right)^2
\end{equation} 
plotted in Fig.~\ref{Fig:Normalized_inertia_contribution} is at most $\mathcal{O}(1)$. This suggests that $\sigma_{\Omega}\ll \sigma_{P}$  in the experiments considered here, that is to say that inertia driven effects are negligible in front of collisionality driven conductivity. Note though that one should, strictly speaking, consider the combined effect of these two contributions, and not each of them separately as done here. Such an analysis requires deriving a generalized model for $\sigma_{\perp}$ combining neutral collisionality and inertia, \emph{i.~e.} Eq.(\ref{Eq:sigma_perdersen}) and Eq.(\ref{Eq:sigma_omega}). This task is left for a future study.

\begin{table*}
\begin{center}
\caption{Derived plasma parameters for selected end-electrodes biasing experiments: electron-ion collision frequency $\nu_{ei}$, electron-neutral collision frequency $\nu_{en}$, ion-neutral collision frequency to ion gyro-frequency ratio $\nu_{in}/\Omega_i$, ion thermal gyro-radius $\rho_i$, parallel conductivity $\sigma_{\parallel}$, Pedersen conductivity $\sigma_P$ and dimensionless parameter $\tau_P=L/a\sqrt{\sigma_P/\sigma_{\parallel}}$. Uncertainties stem from the uncertainties in plasma parameters (primarily the neutral density $n$) given in Table~\ref{Table:Tab1}.}
\label{Table:Tab2}
\begin{tabular}{c c c c c c c c}
\hline
\hline
 & $\nu_{ei}$ [MHz] & $\nu_{en}$ [MHz] & $\nu_{in}/\Omega_i ~[\times10^{-2}]$ & $\rho_i$ [mm] & $\sigma_{\parallel}$ [$\Omega^{-1}$.cm$^{-1}$] & $\sigma_{P}$ [$\times10^{-3}$~$\Omega^{-1}$.cm$^{-1}$] & $\tau_P ~[\times10^{-2}]$\\
 \hline
Q-WVU & $0.52$ & $0.015$ & $0.015$ & $2.7$ & $0.53$ & $1.6\times10^{-6}$ & $0.15$\\
H-KU & $0.049$ & $3.7$ & $1.8 \pm 0.5$ & $6.0 \pm 1.6$ & $0.74$ & $(2.8 \pm 0.8)\times10^{-3}$ & $1.5 \pm 0.2$\\
M-TU & $1.2$ & $1.2$ & $0.51 \pm 0.06$ & $5.4 \pm 0.7$ & $34$ & $(1.62 \pm 0.2)\times10^{-2}$ & $1.1 \pm 0.1$\\
M-UT & $(4.1 \pm 2.5)\times10^{-2}$ & $1.05 \pm 0.17$ & $0.017 \pm 0.002$ & $9.2 \pm 1.1$ & $\left(1.45 \pm 0.35\right) \times 10^{3}$ & $(5.1 \pm 1.8)\times10^{-4}$ & $0.3 \pm 0.02$\\
L-UCLA & $9.3$ & $1.1$ & $0.54$ & $2.9$ & $5.4$ & $0.017$ & $15$\\
M-UW & $1.27 \pm 0.27$ & $7 \pm 7$ & $4 \pm 4$ & $13$ & $\left(7 \pm 6\right) \times 10^{1}$ & $2.0 \pm 2.0$ & $32 \pm 29$\\
H-UNM & $58 \pm 22$ & $44 \pm 23$ & $34 \pm 17$ & $9.3$ & $28 \pm 7$ & $\left(1.1 \pm 0.4\right) \times 10^{2}$ & $320 \pm 100$\\
H-PPPL & $42 \pm 25$ & $123 \pm 35$ & $66 \pm 33$ & $7.1 \pm 3.3$ & $17.1 \pm 2.6$ & $\left(1.1 \pm 0.6\right) \times 10^{2}$ & $52 \pm 17$\\
F-TAE & $0.66$ & $0.24 \pm 0.23$ & $0.13 \pm 0.13$ & $46$ & $\left(1.26 \pm 0.33\right) \times 10^{4}$ & $0.8 \pm 0.8$ & $0.8 \pm 0.5$\\
M-USTC & $0.25 \pm 0.20$ & $22 \pm 5$ & $30 \pm 10$ & $2.1 \pm 0.7$ & $1.3 \pm 0.7$ & $0.16 \pm 0.13$ & $85 \pm 30$\\
\hline
\hline
\end{tabular}
\end{center}
\end{table*}

\section{Summary and Discussion}
\label{Sec:SecIV}

Controlling plasma rotation in magnetized plasmas holds promise for various important plasma applications, ranging from magnetic confinement fusion to plasma separation. One conceptual solution to drive the required rotation is to take advantage of the drift motion arising in crossed-field geometry. Practically though, the important question of how large a perpendicular electric field can be, and over which plasma parameters this field can be obtained, remains uncertain. 

Some elements of response are obtained here by considering the distribution of electric potential along and across a magnetized plasma column characterized by uniform parallel and perpendicular conductivities $\sigma_{\parallel}$ and $\sigma_{\perp}$. Beyond the well known asymptotic regime $\mu^{-1}=\sigma_{\parallel}/\sigma_{\perp}\rightarrow\infty$ for which magnetic field lines are iso-potential, the spatial distribution of potential is shown to be governed by a dimensionless parameter function of the aspect ratio of the plasma column $L/a$ and $\sqrt{\mu}$. For $L/a\sqrt{\mu}\ll1$, the assumption of iso-potential magnetic field lines holds. On the other hand, the potential shows important variations along a given magnetic field line when $L/a\sqrt{\mu}=\mathcal{O}(1)$. 

A look at a selection of past end-electrodes biasing experiments suggests that the inability to control the radial potential profile in some experiments may be attributable to an excessively large ion-neutral collisionality driven perpendicular conductivity $\sigma_P$. Plasma parameters inferred in these experiments indeed correspond to $L/a\sqrt{\sigma_{P}/\sigma_{\parallel}}=\mathcal{O}(1)$. In contrast, biasing experiments where appreciable control over $E_{\perp}$ are found to have $L/a\sqrt{\sigma_{P}/\sigma_{\parallel}}\ll1$. 

While the partial correlation between the simple model developed here and experiments is encouraging, it offers at best so far a tool to identify regimes where electrode biasing is predicted to be ineffective to control the perpendicular electric field.  Turning this model into a predictive model will require capturing all possible contribution to $\sigma_P$, as well as any possible interaction between these contributions. Unfortunately such a comprehensive picture of $\sigma_{\perp}$ is still lacking to date. It will also require accounting for spatial variations in plasma parameters. However, comparison with experimental data will become more and more challenging as these models have more and more free parameters. This is particularly true when relying on past experiments for which one has access to only a limited number of plasma parameters and data points. A recurring shortcoming of available experimental data is the absence of reliable neutral density measurements. To the extent that it is a key input to infer collisionality driven perpendicular conductivity, experiments allowing for detailed mapping of the neutral density appear desirable. Finally, detailed comparison with experiments will require extending this work to include sheath effects at the electrodes.

\section*{Acknowledgments}

The authors would like to thank Dr. S. J. Zweben, I. E. Ochs, E. J. Kolmes and M. E. Mlodik for constructive discussions. This work was supported, in part, by NSF grant PHY-1805316

\appendix

\section{Potential distribution for a single disk electrode}
\label{Sec:appendixA}

\begin{figure*}
\subfigure[~$\sigma_{\perp}/\sigma_{\parallel} = 10^{-2}$]{\includegraphics[]{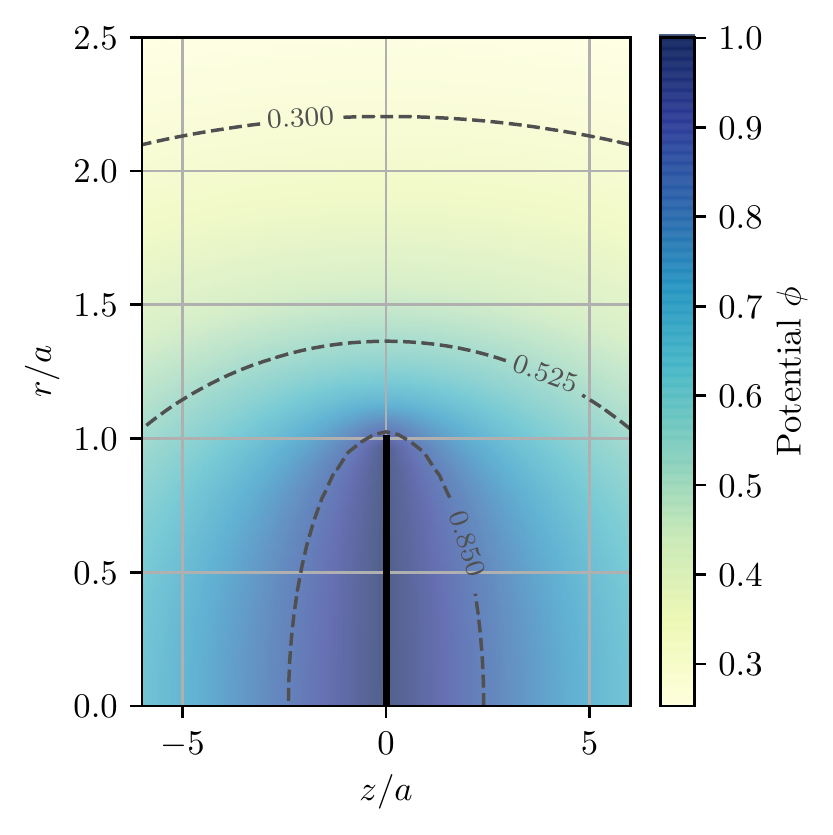}}\subfigure[~$\sigma_{\perp}/\sigma_{\parallel} = 10^{-3}$]{\includegraphics[]{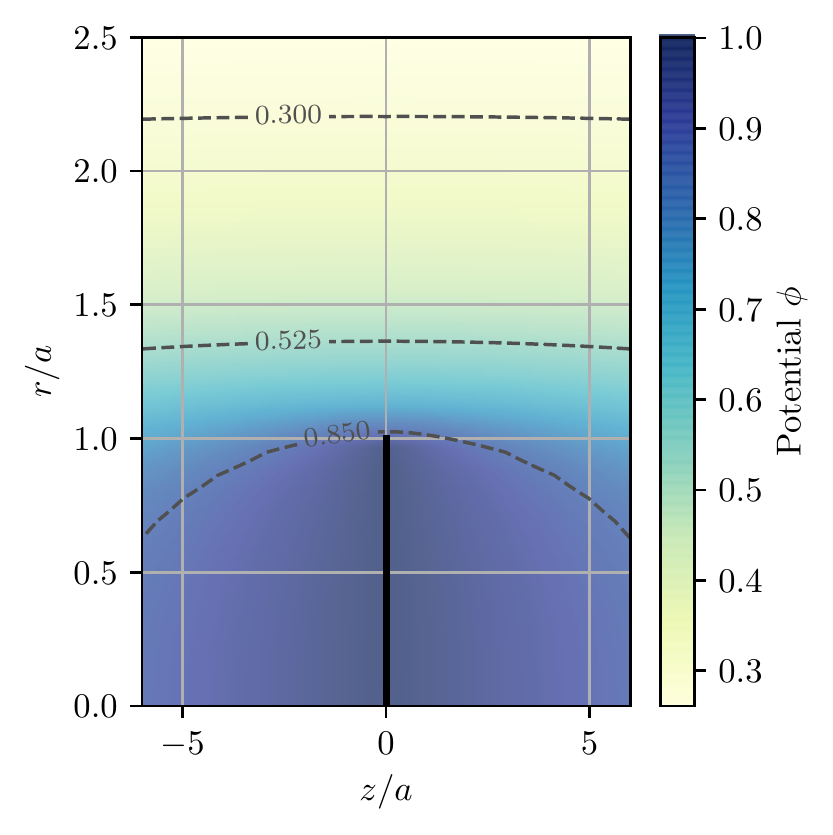}}
\caption{Potential distribution $\phi(r,z)$ obtained from Eq.~(\ref{Eq:Solution_single_electrode}) for an infinitely thin disk electrode of radius $a=1$ at $z=0$ (in solid black) and different values of $\sigma_{\perp}/\sigma_{\parallel}$. The electrode potential $\psi_0=1$. Grey dashed-lines are iso-potential lines.}
\label{Fig:Fig1}
\end{figure*}

Consider a single disk electrode of radius $a$ at potential $\psi_0$ in the $z=0$ plane, as originally treated in vacuum by Weber~\cite{Weber1873}. The potential must vanish when $|z|\rightarrow\infty$. One can thus write
\begin{equation}
\tilde{\phi}(k,z) = A(k)\exp[- k\sqrt{\mu}|z|].
\end{equation}
The boundary conditions at $z=0$ then read
\begin{subequations}
\begin{align}
\phi(r,0)  = \psi_0 \quad &\textrm{for} \quad r<a\label{Eq:BC_pot}\\
\frac{\partial \phi}{\partial z}(r,0)  = 0 \quad &\textrm{for} \quad r>a.\label{Eq:BC_sym}
\end{align}
\label{Eq:BC}
\end{subequations}
Here Eq.~(\ref{Eq:BC_sym}) stems from symmetry. Observing that 
\begin{subequations}
\begin{align}
\int_0^{\infty} j_0(ka)J_0(kr) dk  = \frac{\pi}{2a} \quad &\textrm{for} \quad r<a\\
\int_0^{\infty} k j_0(ka)J_0(kr) dk  =  0 \quad &\textrm{for} \quad r>a,
\end{align}
\end{subequations}
a solution for $A(k)$ satisfying Eq.~(\ref{Eq:BC}) is
\begin{equation}
A(k) = \frac{2 a \psi_0}{\pi k}j_0(ka),
\end{equation}
with $j_0$ the zero$^\textrm{th}$ order spherical Bessel function. The potential distribution in an infinitely long plasma ($L\gg a$) due to the biased electrode at $z=0$ is hence
\begin{equation}
\phi(r,z) = \frac{2\psi_0}{\pi}\int_0^{\infty} \frac{\sin(ka)}{k}J_0(k r) \exp\left(-k\sqrt{\mu}|z|\right) dk.
\label{Eq:Solution_single_electrode}
\end{equation}
The solution $\phi(r,z)$ is plotted for two different values of $\mu = \sigma_{\perp}/\sigma_{\parallel}$ in Fig.~\ref{Fig:Fig1}. As expected, the smaller $\mu$, the smaller the deviation between iso-potential and magnetic field lines. Using Eq.~(\ref{Eq:Int_JO}), Eq.~(\ref{Eq:Solution_single_electrode}) can be integrated in $k$ space to finally yield
\begin{widetext}
\begin{equation}
\phi(r,z) = \frac{2\psi_0}{\pi} \arcsin\left(\frac{2a}{\sqrt{\mu z^2+(a+r)^2}+\sqrt{\mu z^2+(a-r)^2}}\right).
\end{equation} 
\end{widetext}

\section{Potential distribution for two facing electrodes with set radial potential profile and grounded chamber}
\label{Sec:appendixB}

Consider two coaxial disks of radius $a$ terminating a cylindrical conducting chamber of
radius $a$ and length $L$. Let us assume that the chamber is grounded, that is 
\begin{equation}
\phi(r=a,z) = 0 \quad \textrm{for} \quad -L/2\leq z \leq L/2
\end{equation}
and that the electric potential imposed on each disk is
\begin{equation}
\phi \left( r,\pm \frac{L}{2}\right) = \phi_{disk}(r) = \psi_0 J_{0}\left( p_{1}\frac{r}{a}\right)  \quad \textrm{for} \quad r\leq a 
\end{equation}
with $p_{1}\sim 2.4$ the first zero of the zero$^\textrm{th}$ order Bessel function of the first kind $J_{0}$. This choice ensures that the potential is continuous at ($a$,$\pm L/2$). This boundary condition $\phi_{disk}(r)$ also approximates closely a parabolic potential profile since 
\begin{equation}
J_{0}\left( p_{1}\frac{r}{a}\right) = 1 - 2.25 \left(\frac{p_1 r}{3a}\right)^2+\mathcal{O}\left[\left(\frac{r}{a}\right)^4\right]
\end{equation}
with the fourth order term accounting for no more than $3\%$ at mid-radius $r=a/2$. 
 
Solving Eq.~(\ref{Eq:Potential}) for this particular set of boundary conditions, one gets
\begin{equation}
\phi \left( r,z\right) =\psi_0 J_{0}\left( p_{1}\frac{r}{a}\right) \frac{\cosh
\left( p_{1}\tau z/L\right)}{\cosh \left(p_{1}\tau/2\right)} 
\end{equation}
where we introduced the dimensionless parameter $\tau = \sqrt{\mu}L/a$ defined in Eq.~(\ref{Eq:shape_factor}). The radial potential profile in the mid-plane $z=0$ then writes
\begin{equation}
\phi \left( r,z=0\right) = \frac{\phi_{disk}(r)}{\cosh \left(p_{1}\tau/2\right)}.
\label{Eq:bessel}
\end{equation}
Eq.~(\ref{Eq:bessel}) shows that the condition for nearly uniform potential along magnetic field lines is here $\tau\ll 2/p_1$. This is, up to a multiplying factor of $\mathcal{O}(1)$, identical to the criteria derived in Sec.~\ref{Sec:rings} when considering symmetrical rings electrodes.

\section*{References}

\end{document}